\documentclass[aip,amsmath,amssymb,reprint]{revtex4-1}

\usepackage[arrowdel]{physics}
\usepackage{graphicx,psfrag}
\usepackage{dcolumn}
\usepackage{bm}
\usepackage[dvipsnames]{xcolor}
\usepackage{amssymb}
\usepackage{ulem} 
\usepackage{mathtools} 
\usepackage{textgreek} 

\usepackage[utf8]{inputenc}
\usepackage[T1]{fontenc}
\usepackage{mathptmx}
\usepackage{etoolbox}

\newcommand{\br}{\mathbf r}

\newcommand{\pmu}{{\mu^\prime}}

\newcommand{\pn}{{n^\prime}}
\newcommand{\pk}{{k^\prime}}
\newcommand{\psigma}{{\sigma^\prime}}

\newcommand{\ci}{\mathfrak i}

\makeatletter
\def\@email#1#2{%
 \endgroup
 \patchcmd{\titleblock@produce}
  {\frontmatter@RRAPformat}
  {\frontmatter@RRAPformat{\produce@RRAP{*#1\href{mailto:#2}{#2}}}\frontmatter@RRAPformat}
  {}{}
}%
\makeatother
\begin{document}

\preprint{AIP/123-QED}

\title[Chirality-controlled spin scattering through quantum interference]{Chirality-controlled spin scattering through quantum interference}
\author{Jan M. van Ruitenbeek}
 \email{ruitenbeek@physics.leidenuniv.nl}
\affiliation{Huygens-Kamerlingh Onnes Laboratory, Leiden University, NL-2333CA Leiden, Netherlands} 
\author{Richard Koryt\'ar}
\affiliation{Department of Condensed Matter Physics, Charles University, 121 16 Praha 2, Czech Republic} 
\author{Ferdinand Evers}
\affiliation{Institute of Theoretical Physics, University of Regensburg, D-93050 Regensburg, Germany} 

\date{\today}

\begin{abstract}
Chirality-induced spin selectivity has been reported in many experiments, but a generally accepted theoretical explanation has not yet been proposed. Here, we introduce a simple model system of a straight cylindrical free-electron wire, containing a helical string of atomic scattering centers, with spin-orbit interaction. The advantage of this simple model is that it allows deriving analytical expressions for the spin scattering rates, such that the origin of the effect can be easily followed. We find that spin-selective scattering can be viewed as resulting from constructive interference of partial waves scattered by the spin-orbit terms. We demonstrate that forward scattering rates are independent of spin, while back scattering is spin dependent over wide windows of energy. Although the model does not represent the full details of electron transmission through chiral molecules, it clearly reveals a mechanism that could operate in chiral systems.
\end{abstract}

\maketitle

\begin{quotation}
The first observations of, what is now known as chirality induced spin selectivity (CISS), were already reported in 1999 by Ray, Ananthavel, Waldeck and Naaman.\cite{Ray1999}  
Since, many papers have appeared that confirm the general picture: transmission of electrons through chiral molecules selectively favors one of the two spin directions, depending on the handedness of the molecule. 
The effect has been found in photo-emission experiments \cite{Ray1999,Gohler2011,Mishra2013,Kettner2018,Abendroth2019} in electron transport experiments, 
either for small numbers of molecules,\cite{Mishra2020,Kiran2016,Kiran2017,Aragones2017,Xie2011} or in cross-bar configurations across a self-assembled monolayer (SAM) of chiral molecules.\cite{Mishra2020,Liu2020,Mathew2014,Al-Bustami2022} 
Spin polarization efficiencies  have been reported to approach even 100\%,\cite{Lu2019,Al-Bustami2022} and the effects are observed under ambient conditions at room temperature. 
Even more surprising are the related observations that the direction of the magnetization of a thin magnetic film can be determined by the handedness of a monolayer of chiral molecules \cite{BenDor2017} and, 
conversely, a magnetic surface selectively binds one out of the two enantiomers in a racemic mixture.\cite{Banerjee-Ghosh2018,RezaSafari2022} 
{\color{black}Recent reviews are given in Refs.~\onlinecite{Waldeck2021,Naaman2019}, and the connection with chiral spin currents in condensed matter systems is made by Yang {\it et al.}\cite{YangSH2021}}
\end{quotation}

The various attempts at capturing these observations in a theoretical description have recently been summarized by  Evers {\it et al.}\cite{Evers2022} 
Although several possible mechanisms have been discussed that lead to spin selectivity, the discrepancy between the observations and the theory is large. 
The main difficulty that theories encounter is the fact that the spin-orbit coupling in molecules composed of only light elements is extremely small, leaving a gap of many orders of magnitude in the size of the effects between theory and experiment.  
%{\tt Talk about correlation/interaction effects without focus on Fransson}
Although claims have been put forward that the effects can be understood from a combination of spin-orbit interaction and inelastic scattering,\cite{Fransson2020,Das2022} 
it remains unclear how the smallness of the spin-orbit interaction can be addressed by including a, presumably small, correction to it.   
Until this question is resolved we take the point of view that currently {\color{black} few of the proposed ideas are}  capable of explaining the observations. {\color{black} A favorable exception is the work by Dalum and Hedeg{\aa}rd,\cite{Dalum2019} where an amplification of spin-orbit interaction was identified associated with level crossing at multiple energies in chiral molecules. }
%{\tt Include paper of WIS: Binghai Yan; also paper of Joe Subtnik}

%\subsection{Three groups of experiments}
%{\tt Do we need a paragraph on phenomenological classification of experiments in a theory paper? Why would such a classification be motivated here - and in what sense is it different as compared to what has been done in our review paper? }
{\color{black} {\it Three groups of experiments: }Following \cite{Evers2022} we } %propose to 
categorize the experimental evidence in three groups, of increasing level of difficulty met in explaining all the observations. 
The first group of experiments involves the detection of a difference in transmission of electrons of opposite spin through chiral molecules, such as the photo-emission experiments in Refs.~\onlinecite{Gohler2011,Mishra2013,Kettner2018}. These experiments detect the spin of the electrons directly. 

The second and largest group of experimental reports considers changes in electrical resistance of a junction or device as a function of either magnetic field or magnetization of a component.\cite{Mishra2020,Kiran2016,Kiran2017,Aragones2017,Xie2011,Mishra2020,Liu2020,Mathew2014,Al-Bustami2022} 
%In this case the measurement involves changes in charge transport, and the information on the spin transmission is obtained through its effect on the charge conductance as a function of either magnetic field or magnetization of a component.. 
%{\tt The following is theory, while the section is about a phenomenological categorization of experiments. Onsager-passage is perhaps a bit out of place, here?}
As pointed out by Yang {\it et al.}, the resistance of such set-ups 
%the difficulty here lies in the fact that a two-terminal device near equilibrium 
is expected to be insensitive to reversal in the magnetization or the magnetic field 
%of the leads (or of an external magnetic field) 
as known from Onsager's relations, based on fundamental limits imposed by time reversal symmetry.\cite{Yang2019,Yang2020,Yang2021} 
Therefore, magnetoresistance reveals itself in the nonlinear regime, which is much more challenging to understand.
%These authors have also pointed towards avenues for resolving this conflict, by considering deviations from linear current-voltage relations in combination with inelastic scattering. 
%The description of these magnetoresistance effects requires microscopic models involving additional elements on top of the spin-dependent transmission of the first group. 
%Even then, some of the experiments appear to produce a 2-terminal magnetoresistance all the way down to zero bias \cite{Liu2020,Al-Bustami2022}, in apparent violation of Onsager's relations.

The third and final group of experiments comprises observations of (near-)equilibrium properties that are controlled by the handedness of the molecules involved, such as the direction of magnetization of a thin ferromagnet,\cite{BenDor2017} 
or the selective adhesion of enantiomers to a magnetized surface.\cite{Banerjee-Ghosh2018,RezaSafari2022} 
Here, the gap between available theoretical ideas and the experimental observations is largest, and only a few sketchy proposals have been put forward.\cite{Evers2022,Wu2021} 

{\it  Outline of this paper: } Here, we want to focus on the first group of experimental observations only, in the hope that clarification of possible mechanisms for those will lead the way to also resolve the more complicated problems involved in the second and third categories of experiments.
Rather than constructing detailed models for describing any specific experiment or chiral molecule, we focus on simple analytically tractable models, which can guide us in our understanding of the principles involved in CISS. 

One such model  has already been introduced by Michaeli and Naaman:\cite{Michaeli2019} free electrons inside a helical tube with quadratic confining potential. 
By including the spin-orbit interaction resulting from the smooth confining potential an anti-crossing gap opens in the energy spectrum, which results in spin-selective transport through the helical tube. 
This is an important result, because it shows that spin-selective transmission can be obtained generically. However, it falls short in explaining the experiments in that it allows for spin-dependent scattering only in a very narrow energy window of a few meV around the anti-crossing.  
It is important to stress that the model does not predict magnetoresistance (needed for explanation of the second class of experiments) because for evaluation of the full current one needs to include a magnetic electrode.\cite{Korytar2022} 

\begin{figure}[!t]
    \centering
    \includegraphics[width=\columnwidth]{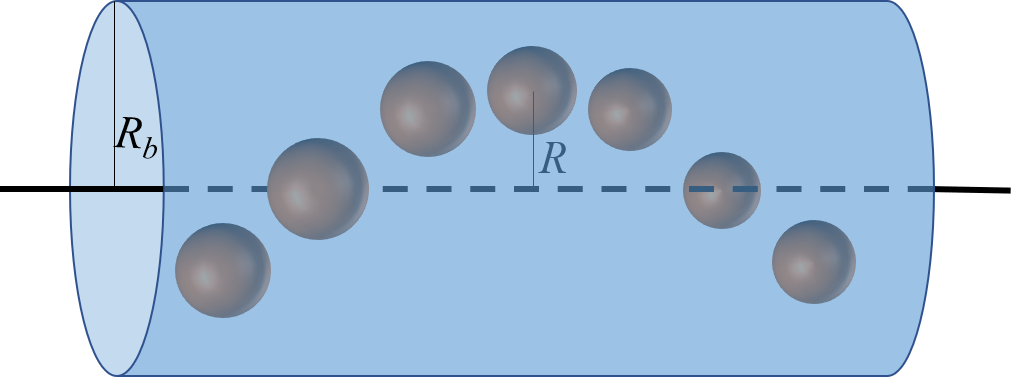}
    \caption{Illustration of the model, which consists of free electrons inside a cylindrical tube, containing a helical string of atomic-like scattering centers.}
    \label{fig:model}
\end{figure}

The model we consider here is that of free electrons traveling inside a straight cylindrical tube, see Fig.~\ref{fig:model}. 
We place an atomic scattering potential off-axis inside this tube, and analyse the spin-dependent scattering due to the spin-orbit interaction at this site by Fermi's golden rule. 
The simplicity of this model permits obtaining analytical expressions, and analyzing the various mechanisms of scattering systematically. 
Chirality is introduced into the problem by arranging a number of atomic scattering potentials as a helical string inside the tube. 
%{\tt the following needs a little work over; Sums over rates are trivially coherent.}
%We find that the effects of this string of atoms can simply be expressed as a coherent sum over scattering 
% rates 
{\color{black} By adding the contributions} 
from the individual atoms
%{\tt The next observation should be made with respect to non-symmorphic structures.}
%We make the important observation 
we find that quantum interference produces large differences in back scattering into the two spin channels, and leads to spin-dependent reflection over wide energy windows. 
The coupling between momentum scattering and spin scattering is seen to result from the non-symmorphic character of the scattering potential.  
% We close by discussing the extend to which the model permits rationalizing the available experimental information, and how it connects to previous theoretical work.

{\color{black}
\section{Model and analytic results}
}

We consider the spin-dependent electronic conduction of chiral molecules as a scattering problem. 
We use a model based on a helical string of atomic-like scattering potentials, sitting inside a perfectly straight wire, carrying Landauer-type conduction channels. The axis of the tube coincides with the $z$-axis in a system of cylindrical coordinates. 
Before introducing the localized scattering potentials, the conductance channels are those for a perfectly smooth and straight cylindrical free-electron-like conductor with hard-wall boundary conditions, and radius $R_b$ (Fig.~\ref{fig:model}). 
The eigenchannels for this perfect wire are given by,
\begin{equation}
    \Psi_{\mu n k}(r,\varphi,z) =  J_\mu(\gamma_{\mu n} r) e^{\ci kz} e^{\ci \mu\varphi} \label{Psi}
\end{equation}
Here, $\gamma_{\mu n}=x_{\mu n}/R_b$, with $x_{\mu n}$ the $n^{\rm th}$ zero of $J_\mu$, the $\mu^{\rm th}$ Bessel function ($n\ge 1$). The energy of this state is
\begin{equation}
    \epsilon_{\mu nk}=\frac{\hbar^2}{2m} \gamma_{\mu n}^2 + \frac{\hbar^2 k^2}{2m}.
    \label{eq:energy}
\end{equation}

The interaction with the atomic-like spin-orbit terms is introduced as a perturbation of these states, by means of Fermi's golden rule.
Although the Coulomb potential of the atoms will be the dominant source of scattering, we propose to focus on the role of the spin-orbit interaction term only, and ignore potential scattering.
Such a model could be a reasonable approximation for a metallic wire, where the electronic states are described by Bloch waves, containing a helical arrangement of heavy ions. 
At this point it is important to stress that we do not aim at a realistic description of a molecular wire. 
Yet, the model will help us to trace similar mechanisms in more realistic models.
The advantage of having an explicit expression for the unperturbed states is that we may evaluate the matrix elements explicitly, which allows us to trace the origin of spin dependent scattering in our model. 

For the spin-orbit interaction term we have,
\begin{align}
    \hat h_0 = \frac{\ci\hbar^2}{4(mc)^2} \boldsymbol{\sigma} \left(
    \frac{\partial v(\br)}{\partial \br} \times \boldsymbol{\nabla}\right). \label{h0}
\end{align}
Here, the differential operator $\partial/\partial_\br$ only acts on the electrostatic potential of the atomic nucleus, $v(\br)$, while the operator $\boldsymbol{\nabla}$ on the right acts on the electron wave function. 
{
%\color{red}
Although the cylindrical confining potential will also produce a contribution to the spin-orbit interaction, we will ignore this in the following, because the symmetry of this potential prohibits any back scattering. Forward scattering cannot lead to spin polarization, as we show in Section I.A.6, below.
}

The spin-flip rates induced by spin-orbit scattering are given by Fermi's golden rule as, 
\begin{align}
    \left.  \frac{1}{\tau_0}\right|_{\mu nk\sigma}  = 2 \pi  
    \sum_{\pmu\, \pn\, \pk, \psigma} &
    |\langle \pmu \pn \pk  \psigma|\hat h_0|\mu n k \sigma\rangle|^2 \, \delta(E_f-E_i).
     %\times p(E_f) 
    \label{FGR} %I have changed label to FGR
\end{align}
%{\tt Give dependency of $E_i$ on ${\bf k}$ etc.. Also, we should drop the occupation numbers; we here work in a single-particle scattering picture, similar to Landauer formalism, which is ignorant of occupation numbers (in lifetimes). }
Here, $E_i = \epsilon_{\mu nk}$ and $E_f = \epsilon_{\pmu \pn \pk}$ are the total energy of the initial and final states.\\

\subsection{Evaluation of the matrix elements}
\subsubsection{Spin-orbit interaction in cylinder coordinates} 
In order to evaluate the matrix elements in (\ref{FGR}) we need to work out the structure of the perturbing Hamiltonian (\ref{h0}). We express the spin operator $\boldsymbol{\sigma}$ in cylindrical coordinates as, 
\begin{align}
\boldsymbol{\sigma}&=\sigma_r \boldsymbol{e}_{r} + \sigma_\varphi \boldsymbol{e}_{\varphi} + \sigma_z \boldsymbol{e}_{z}  \\ 
&= (e^{-\ci\varphi}\sigma_{+} + e^{+\ci\varphi}\sigma_{-})\boldsymbol{e}_{r}
- \ci(e^{-\ci\varphi}\sigma_{+} - e^{+\ci\varphi}\sigma_{-})\boldsymbol{e}_{\varphi}
+ \sigma_z \boldsymbol{e}_{z}, \nonumber
\end{align}
where {\color{black} $\boldsymbol{e}_{r}$, $\boldsymbol{e}_{\varphi}$, and $\boldsymbol{e}_{z}$ are the unit vectors in cylindrical coordinates and,  } 
\begin{equation}
    \sigma_{+} = \hbar\begin{pmatrix}
	0  &  1       \\
	0  &  0 
	\end{pmatrix}, \qquad 
	\sigma_{-} = \hbar\begin{pmatrix}
	0  &  0       \\
	1  &  0 
	\end{pmatrix} = \sigma_{+}^\dagger.  
\end{equation} 
Using coordinates $(r,\varphi,z)$ also for the electrons we can work out the vector products in the expression for $\hat h_0$ as,
\begin{align}
    \boldsymbol{\sigma} \left(
    \frac{\partial v(\br)}{\partial\br} \times \boldsymbol{\nabla} \right)= \hat h_z\sigma_z + e^{-\ci\varphi}\hat h_{+}\sigma_{+} +e^{\ci\varphi}\hat h_{-} \sigma_{-} 
    \label{e10} 
    \end{align}
where we have introduced the real-space operators  
    \begin{align} 
    \hat h_z(r,\varphi,z)&\coloneqq \frac{1}{r} \frac{\partial v}{\partial r} \frac{\partial}{\partial \varphi} - \frac{1}{r}\frac{\partial v}{\partial \varphi} \frac{\partial}{\partial r}  \nonumber \\
    \hat h_{+}(r,\varphi,z) &\coloneqq \frac{1}{r}\frac{\partial v}{\partial \varphi} \frac{\partial}{\partial z}
    - \frac{1}{r}\frac{\partial v}{\partial z} \frac{\partial}{\partial \varphi}  
    - \frac{1}{\ci}\left(\frac{\partial v}{\partial r}\frac{\partial}{\partial z} - \frac{\partial v}{\partial z}\frac{\partial}{\partial r}\right)
    \nonumber 
\end{align}
and $\hat h_{-}\coloneqq \hat h_{+}^*$. 

\subsubsection{General structure of matrix elements} 
Due to \eqref{e10}, two kinds of matrix element appear in Eq. \eqref{FGR}, spin flipping and non-flipping. The spin-conserving one is given by  
\begin{align} 
\langle \pmu \pn \pk  \psigma|\hat h_z\sigma_z|\mu n k \sigma\rangle = 
\langle \pmu \pn \pk|\hat h_z|\mu n k\rangle  \langle\psigma|\sigma_z|\sigma\rangle 
\label{e12} 
\end{align} 
The second matrix element on the rhs is trivially evaluated as 
\begin{align}
    \langle\psigma|\sigma_z|\sigma\rangle = \text{sign}({\sigma}) \delta_{\psigma\sigma}, 
\end{align}
while the first matrix element on the rhs of \eqref{e12} takes the explicit form 
\begin{align} 
    I_z^{\pmu\pn\pk,\mu n k} = \int_0^{R_b} & dr\int_{-\infty}^{+\infty}dz\int_0^{2\pi}rd\varphi 
    \  e^{-\ci\pmu\varphi}  \ e^{-\ci\pk z} \nonumber\\
    \times J_{\pmu}(\gamma_{\pmu \pn}r) \  & \hat h_z(r,\varphi,z) \  e^{\ci\mu\varphi}  \ e^{\ci k z}  J_{\mu}(\gamma_{\mu n}r).
     \label{eq:Iz}
\end{align}
The spin-non-conserving matrix elements are given by  
\begin{align} 
\langle \pmu \pn \pk  \psigma| & e^{\mp \ci \varphi}\hat h_{\pm}\sigma_{\pm}|\mu n k \sigma\rangle =  
\nonumber \\ 
& \langle \pmu \pn \pk|e^{\mp \ci \varphi} \hat h_{\pm}|\mu n k\rangle \langle\psigma|\sigma_{\pm}|\sigma\rangle 
\end{align} 
The second matrix element on the rhs for $\sigma_{+}$ and $\sigma_{-}$ is non-vanishing only for $\sigma=\downarrow$ and $\sigma=\uparrow$, respectively:
\begin{align}
    \langle\psigma|\sigma_{+}|\sigma\rangle & = \delta_{\psigma\uparrow}  \delta_{\sigma\downarrow}, \nonumber \\ \langle\psigma|\sigma_{-}|\sigma\rangle & = \delta_{\psigma\downarrow}  \delta_{\sigma\uparrow}, 
\end{align}
while the first matrix element takes the explicit form 
\begin{align}
    I_\pm^{\pmu\pn\pk,\mu n k} = \int_0^{R_b} & dr\int_{-\infty}^{+\infty}dz\int_0^{2\pi}rd\varphi 
    \ e^{-\ci(\pmu \pm 1)\varphi} \  e^{-\ci \pk z}  \nonumber\\
     \times J_{\pmu}(\gamma_{\pmu \pn}r) \  &  \hat h_\pm(r,\varphi,z) \ e^{\ci\mu\varphi} \ e^{\ci k z} \  J_{\mu}(\gamma_{\mu n}r). 
     \label{eq:Ipm}
\end{align}

Based on these expressions we introduce rates for spin-conserving and spin-flipping processes
%{\tt again, occupation numbers should be dropped, here}
\begin{align}
    \left.  \frac{1}{\tau_z}\right|_{\mu nk\sigma}   {=} 2\pi\xi^2    \sum_{\pmu\, \pn\, \pk} &
     |I^{\pmu\pn\pk,\mu n k}_z|^2 % p(E_f) 
     \delta(E_f-E_i)\nonumber \\ 
    \left.  \frac{1}{\tau_{+}}\right|_{\mu nk\downarrow}  {=} 2\pi\xi^2     \sum_{\pmu\, \pn\, \pk}  &  
      |I^{\pmu\pn\pk,\mu n k}_+|^2 % p(E_f) 
      \delta(E_f{-}E_i)\nonumber \\
    \left.  \frac{1}{\tau_{-}}\right|_{\mu nk\uparrow}  {=} 2\pi\xi^2  \sum_{\pmu\, \pn\, \pk} & 
    |I^{\pmu\pn\pk,\mu n k}_-|^2 %p(E_f) 
    \delta(E_f{-}E_i),
    \label{eq:tau} 
\end{align}
with $\xi = \hbar^2/(4 m^2 c^2)$.
{
%\color{red}
These three scattering rates follow directly from the three terms in Eq.~(\ref{e10}), and can be considered separately because the processes do not mutually interfere.
}
The most interesting situations arise when the rates for up- or down-conversion of the spins, i.e. the lower two lines of \eqref{eq:tau}, differ. 

\subsubsection{Evaluation of the integrals}

The integral for the spatial coordinates in (\ref{eq:Iz}) has two terms inherited from the structure of $\hat h_{z}$. 
The partial derivative $\frac{\partial}{\partial \varphi}$ at the right in the operator $\hat h_{z}$, acting on a wave function $|\mu n k \sigma\rangle$, produces $\ci\mu$.
The two terms can be combined through partial integration. 
For the first term we use for the integral over $r$, 
\begin{align}  
    \ci  \mu \int_{0}^{R_b}  & \, \frac{\partial v}{\partial r}\,  J_{\pmu} J_{\mu}  \, dr =  \\
   & - \ci \mu  \int_{0}^{R_b} \,v\,\left(  J_{\mu} \frac{dJ_{\pmu}}{d r} + J_{\pmu} \frac{dJ_{\mu}}{dr}  \right)\, dr. \nonumber
\end{align}
where we write $J_{\mu}$ and $J_{\pmu}$ as short for $J_{\mu}(\gamma_{\mu n} r)$ and $J_{\pmu}(\gamma_{\pmu \pn} r)$.

For the second term we use,
\begin{align} 
    \int_{0}^{2\pi} \, e^{-\ci (\mu -\pmu )\varphi} \frac{\partial v}{\partial\varphi} \, & d\varphi = \\
    & - \ci (\mu -\pmu ) \int_{0}^{2\pi} v\, e^{-\ci (\mu -\pmu ) } \, d\varphi. \nonumber
\end{align}

The two terms combine into,
\begin{align} \label{eq:I_z}
    I_z^{\pmu\pn\pk,\mu n k} &=  -\ci \int_0^{R_b}dr\int_{-\infty}^{+\infty}dz \,\int_0^{2\pi}d\varphi\ \, v(\mathbf{r}) \, e^{\ci(\mu -\pmu) \varphi} 
    \nonumber\\
    & \times e^{\ci (k- \pk) z}  \, \left(  \mu  J_{\mu} \frac{dJ_{\pmu}}{d r} + \pmu J_{\pmu} \frac{dJ_{\mu}}{d r}  \right) 
\end{align}
% This integral vanishes when $\mu=-\pmu$ and $n=\pn$.

By similar steps we obtain for the corresponding integrals for the operators $\hat h_{\pm}$,
\begin{align} \label{eq:I_pm}
    I_\pm^{\pmu\pn\pk,\mu n k} &=  \int_0^{R_b}dr\int_{-\infty}^{+\infty}dz\int_0^{2\pi}d\varphi \, v(\mathbf{r}) \,  e^{\ci(\mu -\pmu \mp 1) \varphi} \times
    \nonumber\\
       e^{\ci (k- \pk) z} &\left[ ( \mu \pk - \pmu k) J_{\pmu} J_{\mu} 
        \pm r \left( k J_{\mu} \frac{dJ_{\pmu}}{d r}  + \pk J_{\pmu} \frac{dJ_{\mu}}{d r} \right) \right]. 
\end{align}
Here, we have assumed that the potential $v(\mathbf{r})$ vanishes for $z \rightarrow \pm\infty$, but otherwise it has not yet been specified.

\subsubsection{Limiting case: a single `atom'}
%{\tt Change wording a bit - "spurious" is perhaps not quite ...}
Note that the potential due to a single atomic-like scatterer does not define a chiral structure. 
Yet, we find a difference between the integrals for $\tau_+$ and $\tau_{-}$ in (\ref{eq:I_pm}) even for a single scattering center. 
This spin-dependent scattering arises from the combination of the finite orbital angular momentum $\hbar\mu$ and $\hbar\pmu$ of the incoming and scattered waves, in combination with the off-axis position of the `atom'. 
Experiments that permit selection of the angular momentum of incoming waves may observe this spin-dependent scattering, but typical experiments select the incoming waves by energy only. 

This %`spurious' 
chiral effect {\color{black} that is not associated with a chiral potential} is removed by summation of contributions of all incoming electrons at the same energy. This can be seen from the fact that $I_{+}^{\pmu\pn\pk,\mu n k} = -I_{-}^{-\pmu\pn\pk,-\mu n k} $, so that $|I_{+}^{\pmu\pn\pk,\mu n k}|^2 = |I_{-}^{-\pmu\pn\pk,-\mu n k}|^2 $.
Therefore, {\color{black} for a single `atom',} 
\begin{align}
    \left.  \frac{1}{\tau_{+}}\right|_{\mu nk\downarrow} = \left.  \frac{1}{\tau_{-}}\right|_{-\mu nk\uparrow}
    \label{eq:equal}
\end{align}
Since the states with quantum number $\mu$ are degenerate with those having $-\mu$ we should consider the sum of the scattering rates for $\mu$ and $-\mu$ and, thus, this sum is equal for $\tau_{+}$ and $\tau_{-}$. 

Concluding, as we should expect, a single atomic-like potential does not represent a chiral structure and %does not produce any spin dependence in scattering. 
{\color{black} only produces spin-dependent scattering when we can conceive of an experiment that selects the orbital states of the electron waves. }
Below we will combine the scattering of a string of `atoms' and demonstrate that spin dependence results from a helical arrangement of single scatterers.

%\subsubsection{A point-like scatterer} 

The potential $v(\br)$ in (\ref{h0}) describes the electrostatic potential due to the `atoms' inside the tube. %Let us first 
{\color{black} As a first step, we} limit this to a single site at $(R,\Phi,Z)$. %We will discuss how the effects of a string of `atoms' add up at a later stage, below. 
We calculate the matrix elements explicitly by adopting a minimal model for the scattering potential $v(\br)$. 
The potential is normally given by the Coulomb interaction between the electrons and the effective core. For the purpose of the minimal model we have the freedom of adjusting the actual form of this potential, and for computational convenience we choose it to be a delta function 
\begin{equation}
    v(r,\varphi,z) =  K_0 \delta(r-R)\frac{1}{r}\delta(\varphi-\Phi) \delta(z-Z) .
\end{equation}
With this, the integrals for the spatial coordinates in matrix elements for $\hat h_{z}$, $\hat h_{+}$ and $\hat h_{-}$ become, 
\begin{align} \label{eq:I-delta}
    &\quad I_{z}^{\pmu\pn\pk,\mu n k} =  -\ci K_0  e^{\ci(\mu - \pmu)\Phi} e^{ \ci (k-\pk) Z }  \times  \\
     &\qquad \qquad \left[ \frac{1}{r}   
    \left( \mu J_{\mu} \frac{dJ_{\pmu}}{dr} + \pmu J_{\pmu} \frac{dJ_{\mu}}{dr}\right) \right]_{r=R} , \nonumber \\
    &\quad I_\pm^{\pmu\pn\pk,\mu n k} =  K_0  e^{\ci(\mu - \pmu \mp 1)\Phi} e^{ \ci (k-\pk) Z } \nonumber \times  \\
     &\qquad \left[ \frac{1}{r} ( \mu \pk  - \pmu k ) J_{\mu} J_{\pmu} \pm  
    \left( k J_{\mu} \frac{dJ_{\pmu}}{dr} + \pk J_{\pmu} \frac{dJ_{\mu}}{dr}\right) \right]_{r=R} . \nonumber
\end{align}
The terms in square brackets depend on the quantum numbers, but otherwise only on the radial distance $R$ for the position of the atomic potential. The dependence on $Z$ and $\Phi$ is completely covered by the phase factors in the first lines of \eqref{eq:I-delta}.

\subsubsection{A helical string of atomic-like scatterers}

In order to implement a chiral structure we arrange $2N+1$ identical `atoms' along a helix inside the cylindrical conductor at positions $(R,\nu \theta, \nu s)$ for $\nu=-N, \ldots, N$, where the constants $\theta$ and $s$ describe the chirality of the `molecule'.
Thus, we write the potential $v$ in the form of a sum over the potentials of the individual nuclei,
\begin{equation} 
    v_{N}(\br) = \sum_{\nu = -N}^N v_0(r - R,\, \varphi-\nu\theta, \, z - \nu s ),
\end{equation}
where $v_0$ is the potential due to a single scattering center. 

In order to evaluate the matrix elements for this potential, we insert it in Eqs.~(\ref{eq:I_z}) and (\ref{eq:I_pm}). The scattering amplitudes from each of the `atoms' in the string receive phase factors according to \eqref{eq:I-delta}, which allows us to express the integrals as,
\begin{align} \label{eq:I_z_nu}
    I_z^{\pmu\pn\pk,\mu n k} &= I_z^{\pmu\pn\pk,\mu n k} (v_0)  
    \sum_{\nu =-N}^N e^{\ci \nu (s \Delta k + \theta\Delta\mu)} \nonumber \\
    = & I_z^{\pmu\pn\pk,\mu n k} (v_0)  \quad
    \frac{\sin{((N+\tfrac{1}{2})(s \Delta k + \theta \Delta \mu))}}{\sin{(\tfrac{1}{2} (s \Delta k + \theta\Delta\mu))}}  \nonumber \\
    I_{\pm}^{\pmu\pn\pk,\mu n k} &=
    I_{\pm}^{\pmu\pn\pk,\mu n k} (v_0)  
    \sum_{\nu =-N}^N e^{\ci \nu (s \Delta k + \theta(\Delta\mu \mp 1))} \nonumber \\
    =  I_{\pm}^{\pmu\pn\pk,\mu n k} & (v_0)  \quad
    \frac{\sin{((N+\tfrac{1}{2})(s \Delta k + \theta (\Delta \mu\mp 1)))}}{\sin{(\tfrac{1}{2} (s \Delta k + \theta(\Delta\mu \mp 1)))}}  ,
\end{align}
where $\Delta k = \pk -k $ and $\Delta \mu =  \mu-\pmu$. The integrals on the rhs are evaluated for the potential $v_0$ due to a single scattering center.
The sums in \eqref{eq:I_z_nu} generally lead to near-cancellations, unless the argument $s (\pk -k) + \theta(\mu-\pmu \mp 1)$ is close to a multiple of $2\pi$. This quantum interference effect strongly selects specific channels for spin scattering.

This is the central result of this paper: quantum interference breaks the symmetry between spin-up and spin-down scattering. 
For example, consider initial and final states with $\pmu = \mu$ and $\pn = n$. 
Energy conservation requires $\pk = \pm k$. For back scattering we have $\pk = -k$, so that the arguments are $x = -2ks - \theta$ for $I_{+}$ and $x = -2ks + \theta$ for $I_{-}$. This breaks the symmetry of \eqref{eq:equal}, 
in particular for the energy range for which $k \simeq \theta/2s$ so that $x$ approaches zero for $I_{-}$, while $x  \simeq 2\theta$ for $I_{+}$, which will be typically far from zero. The other spin channel is selected when $x = -2ks - \theta \simeq -2\pi$.

At higher energy many combinations of quantum numbers lead to constructive interference, and in order to evaluate this we need to resort to numerical evaluation of the expressions.

\subsubsection{Absence of spin-dependent forward scattering}

In order to suppress %spurious chirality
{\color{black} scattering that is not associated with the chiral shape of the potential }  we consider the average of scattering of all states available at a given energy,
\begin{align}
    \left.  \frac{1}{\tau_{\pm}}\right|_{E}  {=}     \frac{1}{n_{\tau}} \sum_{\mu\, n}  \sum_{k>0}  
    \left.  \frac{1}{\tau_{\pm}}\right|_{\mu n k\sigma} \delta(E-\epsilon_{\mu n k}),
     \label{eq:tau-E} 
\end{align}
where the energy for each of the states $\epsilon_{\mu n k}$ is given by \eqref{eq:energy}, and $n_{\tau}$ counts the numbers of states available at this energy.
The polarization of the spin scattering can then be defined as,
\begin{equation}
    P(E) = \frac{ \left.  \frac{1}{\tau_{+}}\right|_{E} - \left.  \frac{1}{\tau_{-}}\right|_{E} }{ \left.  \frac{1}{\tau_{+}}\right|_{E} + \left.  \frac{1}{\tau_{-}}\right|_{E}  + \left. \frac{1}{\tau_{z}}\right|_{E} }
    \label{eq:ratio}
\end{equation}

Remarkably, we find that the polarization for forward scattering ($\pk>0$) is identical to zero. In order to demonstrate this let us consider incoming states with energy $E$ and quantum numbers $\mu,n,k>0$.
The spin-flip scattering rates averaged over states at energy $E$ can be written as,
\begin{subequations}
\label{seq1}
\begin{align}
\label{seq1a}
\left.\frac{1}{\tau_+} \right|_E^\text{fwd.} &=  \frac{2\pi\xi^2}{n_{\tau}} \, \sum_{\mu n k}\sum_{\pmu \pn \pk}\
\left|\bra{\pmu \pn \pk \downarrow}\, \hat h_0\, \ket{\mu n k\uparrow}\right |^2 \nonumber \\
&\qquad\qquad\times\delta(E-\epsilon_{\mu'n'k'})
\delta(E - \epsilon_{\mu nk}) \\
\label{seq1b}
\left.\frac{1}{\tau_-} \right|_E^\text{fwd.} &= \frac{2\pi\xi^2}{n_{\tau}} \, \sum_{\mu n k}\sum_{\pmu \pn \pk}\
\left|\bra{\pmu \pn \pk \uparrow}\, \hat h_0\, \ket{\mu n k\downarrow}\right |^2 \nonumber \\
&\qquad\qquad\times\delta(E-\epsilon_{\mu'n'k'})
\delta(E - \epsilon_{\mu nk}).
\end{align}
\end{subequations}
In order to account for forward scattering only we restrict the final-state momenta to positive values, $k'>0$.
In this case, the indices $\mu, n, k$ and $\pmu,\pn,\pk$ run through identical sets of quantum numbers
in the \eqref{seq1}. When one uses the fact that $\hat h_0$ in \eqref{seq1} is hermitian, and relabels the
indices, it it easy to see that
\begin{equation}
\left.\frac \hbar{\tau_+} \right|_E^\text{fwd.} =
\left.\frac \hbar{\tau_-} \right|_E^\text{fwd.}.
\end{equation}

This identity does not apply for back scattering, because in this case the summation over $k$ and $\pk$ cannot be interchanged. In the following we will focus on the properties of back scattered electron spins.

\section{Numerical evaluation}

\begin{figure}[b!]
    \centering
    \includegraphics[width=\columnwidth]{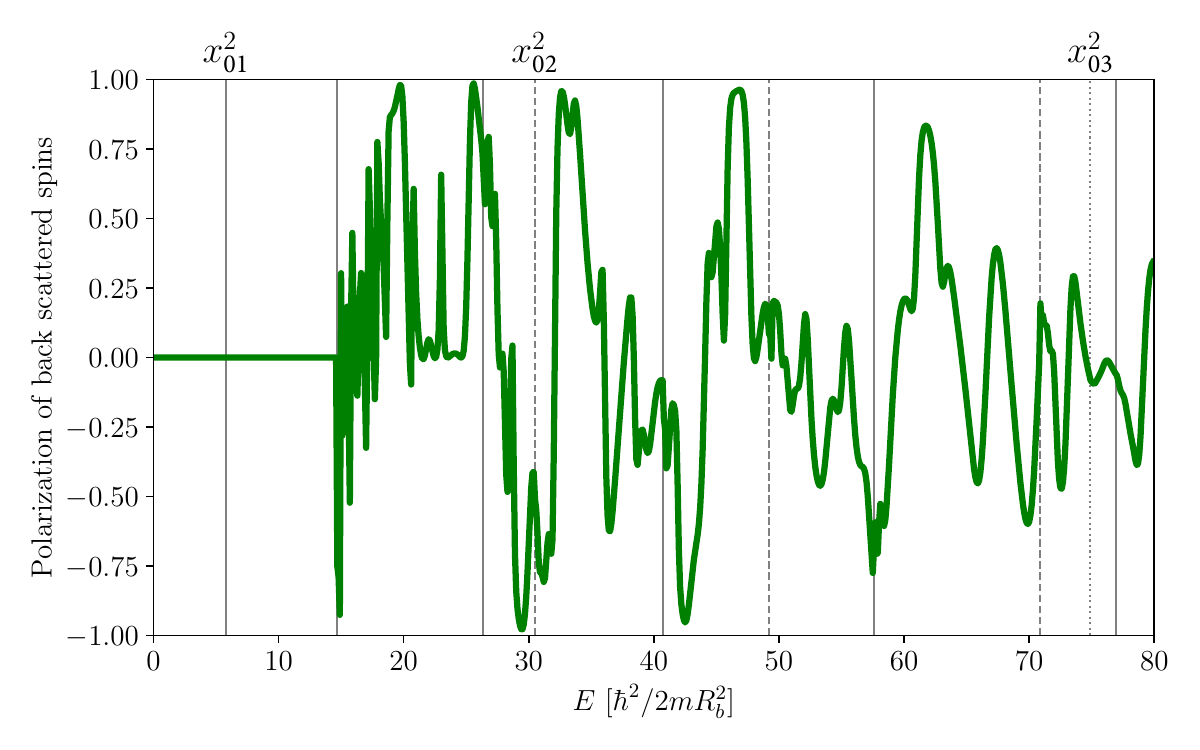}
    \caption{Constructive interference of the spin scattering in a string of 21 `atoms' (N=10). 
    Here $P$ is the ratio of the rates for back scattering spins from up to down and the reverse process, as given by Eq.~(\ref{eq:ratio}), as a function of energy $E$, in units $\hbar^2/(2 m R_b^2)$. 
    The angle between successive scattering centers in the helical string is set at $\theta = 2\pi/10$, and the distance between the sites $s=0.85$ in units of $R_b$. The vertical lines mark the points of opening of new conductance channels for $n=1$ (solid), $n=2$ (dashed), and $n=3$ (dotted). Successive lines in the same style have increasing values of $\mu$.
    Changing the sign of the chirality produces a curve with the opposite sign.}
    %{\color{red} change labels $x_{00}^2 \rightarrow x_{01}^2$ etc.}
    \label{fig:ratios}
\end{figure}

The many contributions to the sums in \eqref{eq:tau} require numerical evaluation. 
%{\color{red} (see also Supplementary Material). }
For back scattering we find the result plotted in Fig.~\ref{fig:ratios}. The important result we obtain is that $P(E)$ is typically large, and even approaches complete polarization in some regions of energy. 
When we change the sign of the chirality (by $s\rightarrow -s$, or $\theta \rightarrow -\theta$) the graph is reflected around the horizontal axis, as expected. We find that the spin-conserving term $1/\tau_{z}$ only makes a minor contribution to the total scattering rates.

Initially, $P(E)$ is zero, because below $(\gamma_{01}/R_b)^2 = 5.78$ there are no conduction channels available. 
After that point the first channel opens, with $\mu=0,\,n=1$, $\pmu=0,\,\pn=1$, and this remains the only channel until $(\gamma_{11}/R_b)^2 = 14.68$. 
In this range of energies the transmission and reflection of spins is exactly balanced, in accordance with Kramers' degeneracy for a single-channel conductor.\cite{Bardarson2008} 
In our expressions, this absence of spin scattering for the lowest conductance channel can be read from \eqref{eq:I-delta}: 
the first term in the square brackets vanishes for $\mu=\pmu=0$, and the second term cancels because the Bessel functions for $\mu$ and $\pmu$ are the same, and because $\pk = -k$ for back scattering into the same channel.

Once we cross $E=14.68$ additional conductance channels become accessible, having non-zero angular momentum quantum numbers $\mu,\, \pmu=\pm 1$, with $n=\pn=1$. 
The contributions of these channels allow for the integrals \eqref{eq:I-delta} to become finite, and the gradual increase of $k$ and $\pk$ above $E=14.68$ leads to rapid oscillations of $P(E)$ due to the phase factors in \eqref{eq:I_z_nu}.  
The polarization continues to fluctuate with energy, but longer-period components take over. 
In some cases it is possible to trace the saturation of $P$ near 1 to a contribution for which the argument of the phase factor in the interference $x=s(k-\pk)+\theta(\mu-\pmu \pm 1)$ becomes very small, so that all terms add constructively. 
When this happens the spin scattering rate for these terms shows a very long period oscillation as a function of energy and as a function of the numbers of `atoms' in the string.

Importantly, we find that $P(E)$ maintains predominantly the same sign over wide ranges of energy. For example, $P$ is almost exclusively positive between $E=17.5$ and $E=27.5$.
Since many experiments do not select a sharply defined energy, but a finite range of energies contributes to the signal, it is useful to integrate the scattering rates over an energy window. 

\begin{figure}[t!]
    \centering
    \includegraphics[width=\columnwidth]{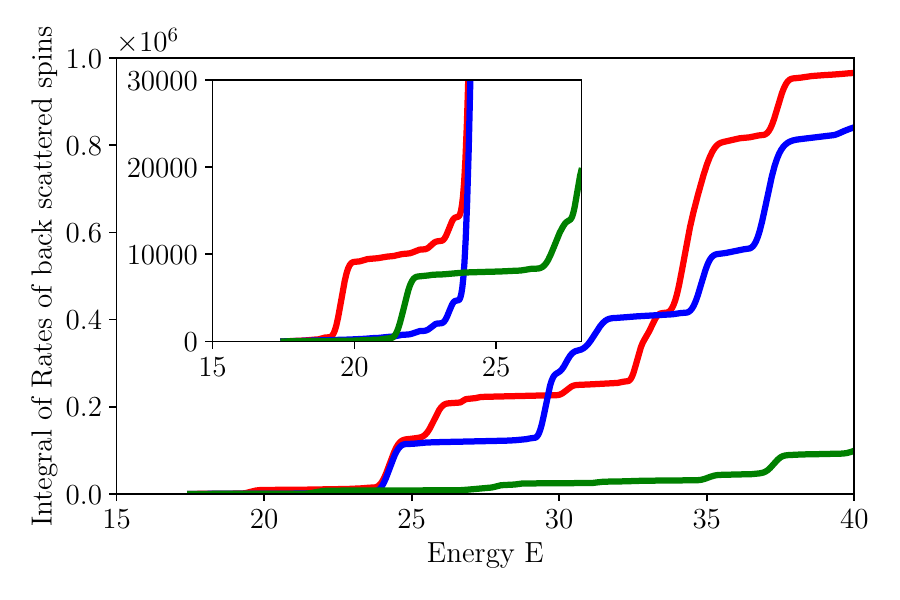}
    \caption{Rates $\left.  1/\tau_{+}\right|_{E}$ (red), $\left. 1/\tau_{-}\right|_{E}$ (blue), and $\left.  1/\tau_{z}\right|_{E}$ (green)  integrated over energy from $E=17.5$, for $2N+1=41$. The rates are given in arbitrary units. The inset shows the low-energy range on an expanded scale.}
    \label{fig:integrated-rates}
\end{figure}

Figure~\ref{fig:integrated-rates} shows the rates $\left.  1/\tau_{+}\right|_{E}$ (red), $\left. 1/\tau_{-}\right|_{E}$ (blue), and $\left. 1/\tau_{z}\right|_{E}$ (green) for $2N+1 = 41$ atomic scattering centers, integrated from E = 17.5 to E.
Clearly the difference between the two spin scattering directions is large, in particular at the lower energy end. At higher energies the predominance of spin-up vs. spin-down scattering alternates.

\begin{figure}[t!]
    \centering
    \includegraphics[width=\columnwidth]{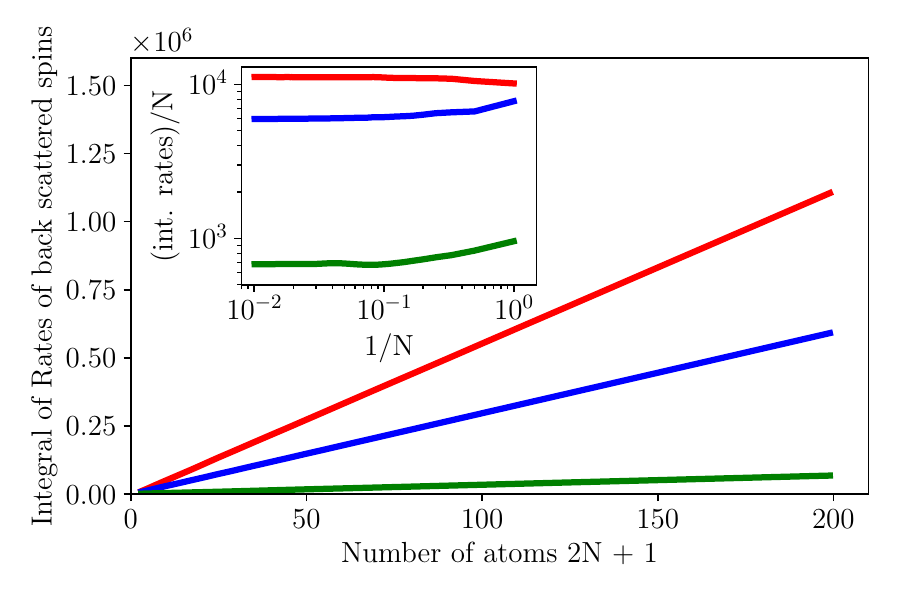}
    \caption{Energy-integrated back scattering rates $\left.  1/\tau_{+}\right|_{E}$ (red), $\left. 1/\tau_{-}\right|_{E}$ (blue), and $\left. 1/\tau_{z}\right|_{E}$ (green) as a function of the length of the helical string. The integration over energy is taken from $E=17.5$ to $E=27.5$. 
    The number of atoms in the string is $2N+1$. At $N=0$ the tube contains one atom, and the blue and red rates are equal, as expected. {\color{black} The inset plots the rates devided by $N$, as a function of $1/N$, which demonstrates the saturation to a constant slope of the rates for large $N$.} The rates are given in arbitrary units.}
    \label{fig:length-dependence}
\end{figure}

One of the hallmark observations in the experiments is a roughly linear dependence of the spin polarization with the length of the molecule. 
In our model, at any given energy we find that the scattering rates oscillate with the number of `atoms' $2N+1$ included in the helical string. 
This oscillation is often rapid, but long-period oscillations are found near the points where $P(E)$ approaches 1 or -1 (Fig.~\ref{fig:ratios}).
However, as noted above, experiments typically measure the signals due to a finite range of energies. 
Figure~\ref{fig:length-dependence} shows the two spin scattering rates, integrated from $E=17.5$ to $E=27.5$, as a function of the length of the helical string, where the number of atoms is $2N+1$.
The observed dependence is close to linear, while the polarization $P$ remains approximately constant. 
The linear dependence of the spin scattering rates persists (at least up to $N=100$), so that the spin scattering can, in principle, dominate other sources of scattering for long molecules.
{
%\color{red} 
The linear dependence will ultimately saturate due to interactions that are not included in our model, such as multiple scattering and inelastic scattering. 
}

\section{ Discussion and conclusion}

The model presented above was investigated in order to trace a possible origin of the chirality-induced spin selectivity. 
We find that quantum interference of partial waves scattered off atomic spin-orbit interactions leads to selective back scattering of one spin component over the other. 
Although this mechanism appears to be intuitively appealing, we cannot claim that we offer a quantitative explanation of the CISS effect. 
The nature of the unperturbed electronic states is idealized in our model, and the spin-orbit interaction at the atomic sites is represented by a delta-function potential, which is clearly not realistic. 
Therefore, the quantitative outcomes for the scattering rates cannot be taken literally for a comparison with experiments. 

The strong points of the mechanism proposed here is that it is robust, conceptually simple, and may be applicable to a wider range of unperturbed molecular wave functions. 
In our model, spin selectivity is found in wide ranges of energy, despite the smallness of the spin-orbit interaction. 
This contrasts with the model proposed by Michaeli,\cite{Michaeli2019} which produces spin selectivity of order unity only in a narrow window of a few meV at low energies.
In our case it is found for all energies above a certain threshold value. 
Furthermore, we find that the spin scattering rates increase almost linearly with the length of the 'molecule', in agreement with observations.\cite{Gohler2011}  

%We have ignored scattering of the electronic states due to the direct Coulomb potential on the atomic sites. 
%In more realistic models such interaction would already be included in the initial wave functions of the conductance channels, in the form of molecular orbitals or Bloch waves. 
%By adding the spin-obit interaction as a perturbation to such electric wave functions we suspect that similar quantum interference terms will surface. {\tt Ferdinand, can we find more arguments to support for this proposition?}

%An indication that the mechanism can be found in more realistic computations is obtained from the work by Dalum and Hedeg{\aa}rd \cite{Dalum2019}. 
%They consider a tight-binding model of a chain of carbon atoms, that is continuously deformed from a straight wire into a helix of variable pitch. 
%They obtain spin-dependent transmission, which oscillates as a function of energy and helical twist angle. 
%They attribute the effect as a result of amplification of the spin-orbit interaction by the appearance of many level crossings, 
%but it would be interesting to investigate whether this could also be viewed as an effect of quantum interference.

An interesting feature of our model is that the sign of the spin scattering is {\em not} uniquely determined by the handedness, as shown by the plot in Fig.~\ref{fig:ratios}, but also by the helicity $\theta/s$ and by the energy range that we consider. 
Most experiments have compared the effects of the sign of chirality by comparing the two enantiomers, or have studied similar molecular structures as a function of length, all under the same experimental conditions. 
In a system that can be described by this quantum interference mechanism one would expect to observe sign changes when varying the helical pitch, even when maintaining the same sign of chirality, or when probing the system in a different energy range.

{
%\color{red}
The model considered here is consistent with Landauer's picture of a phase-coherent scattering problem. It transitions to a classical resistance only when we add inelastic scattering to the description, and consider the limit of very long helical wires. Even at room temperature, for most chiral molecules probed in experiments the inelastic scattering length is much longer than the length of the molecule. On the other hand, the long DNA strands tested by Gohler {\it et al.}\cite{Gohler2011} are possibly long enough for temperature-induced dephasing to become observable.
}

In conclusion, we propose a simple model of constructive quantum interference as a mechanism giving rise to spin-selective electron reflection.
The mechanism proposed here may guide the design of experiments and the analysis of more realistic computations.
We have limited our discussions to possible explanations of spin-selectivity in the transmission properties of chiral molecules, and the model presented here offers a simple and intuitive mechanism that may be transferable to actual molecular systems.
We have refrained from touching upon experiments that involve charge detection, rather than spin directly, because of the additional complications involved in describing the spin-to-charge conversion (both in terms of the modeling, and regarding the proper design of the experimental conditions).
The observation of the third class of experiments, revolving around near-equilibrium properties of enantiomer absorption on magnetized surfaces, pose even greater difficulties for explanation, and we have not attempted at addressing those.
A proper understanding of spin-selective transmission may form a solid basis for proceeding with developing an explanation for the second two classes of experiments.

\begin{acknowledgments}
FE acknowledges support by the German Science Foundation. The work by JMvR is part of the research program of the Netherlands Organisation for Fundamental Research, NWO. RK acknowledges the Czech Science Foundation (project no. 22-22419S).
We gratefully acknowledge lively discussions with Per Hedeg{\aa}rd, J\c{e}drzej Tepper, Sense Jan van der Molen, Peter Neu, Tjerk Oosterkamp, and Julian Skolaut.
\end{acknowledgments}

\nocite{*}
\bibliographystyle{apsrev4-2}
\bibliography{literature.bib}

\end{document}